\documentstyle[11pt,epsfig]{article}
\newcommand{\ket}[1]{| {#1} \rangle}
\newcommand{\bra}[1]{\langle {#1} |}
\newcommand{\ave}[1]{\langle {#1} \rangle}

\begin{document}

\begin{titlepage}

\begin {center}
{\large \bf  Single Boson Images Via an Extended 
Holstein-Primakoff  Mapping}\\
\vspace*{0.5cm}
Zoheir Aouissat\\

Institut f\"{u}r Kernphysik, Technische Universit\"at 
Darmstadt, Schlo{\ss}gartenstra{\ss}e 9,\\ D-64289 Darmstadt, Germany
\end{center} 
\vspace*{2.cm}
\begin{abstract}
%=====================================================================
The Holstein-Primakoff mapping for pairs of bosons is extended in order
to accommodate single boson mapping.  
The proposed extension allows a variety of applications and especially
puts the formalism at finite temperature on firm grounds. 
The new mapping is applied to the  $O(N+1)$ anharmonic oscillator with
global symmetry broken down to $O(N)$. 
It is explicitly demonstrated that $N$-Goldstone modes appear.
This result generalizes the  Holstein-Primakoff 
mapping for interacting bosons  as developed in ref. \cite{ASW}.  

%=====================================================================
\end{abstract}

\vspace*{2.cm}


{\bf Report:} IKDA 1999/27

\end{titlepage}
\newpage

\newpage
%%%%%%%%%%%%%%%%%%%%%%%%%%%%%%%%%%%%%%%%%%%%%%%%%%%%%%%%%%%%%%%%%%%%%%%
%
\section{Introduction}
%
%%%%%%%%%%%%%%%%%%%%%%%%%%%%%%%%%%%%%%%%%%%%%%%%%%%%%%%%%%%%%%%%%%%%%%%
Boson expansion theory (BET) has played a significant role over the past
decades in our understanding of the nuclear many-body problem.
Starting with the pioneering work of Marumori and co-workers \cite{MARU}, 
and of Belyaev and Zelevinsky \cite{BEZE},  
the interest in this subject has culminated in the eighties through 
the formulation of the interacting boson model \cite{IAAR}. 
Of particular interest, not only
for the many-body problem, but also, as it has become apparent, 
for quantum-field theory (QFT),
is the perturbative boson expansion (PBE) approach.
Extensive use  of it has been  made 
in nuclear physics, in order to extract   
anharmonicities beyond the Random-Phase Approximation (RPA) 
(see ref.\cite{RS80,KLMA} for  reviews).
Up until very recently its application to QFT has not attracted much attention and,
therefore, has not been fully developed so far. 
The Holstein-Primakoff mapping for boson pairs, first introduced in  
\cite{BPD,CDDF}, was recently applied, however,  to the $O(N)$  
vector model \cite{ASW}. 
It was demonstrated that the mapping is able to 
systematically classify the dynamics according to the $1/N$-expansion,  
rendering a promising and efficient alternative to the well-known
functional methods. Furthermore, considering the model in the phase of 
spontaneously broken symmetry, the powerful machinery of the PBE approach 
as developed for deformed nuclear systems could be transcribed to QFT. 
As a consequence the Goldstone theorem as well as the whole hierarchy of Ward
identities were exactly satisfied \cite{ASW}. \\
However, the PBE in general, and the Holstein-Primakoff mapping (HPM) in particular 
\cite{HOPR}, rely on the bosonisation of pairs of particles. Thereby, 
images for particle pairs are generated in an ideal Fock-space, while
single particle images are absent after mapping.    
This problem has been appreciated for the fermionic case in the early days 
of the boson expansion theory. 
Marshalek has proposed an extension of HPM for fermions 
in order to allow for a perturbative boson expansion for both even and odd 
nuclei \cite{MAR}. In the present letter, we point out the occurrence 
of the same problem in the case of the PBE for purely bosonic models. \\  
The need for an extended bosonic HPM to include single bosons
clearly revealed itself in \cite{ASW} where the lack of ideal single-boson states 
was an obstacle for defining unambiguously the two-point function for the
Goldstone mode. While to leading order in the $1/N$-expansion this problem 
was circumvented in \cite{ASW}, 
a next-to-leading order calculation makes an extended 
version of the HPM mandatory.\\
Finite-temperature applications of the PBE approach is another issue
where an extended HPM to include single bosons is definitely called for.   
In the following, we wish to sketch a derivation of an extended
version of HPM  for bosons.   
We will also discuss  an application to the $O(N+1)$ anharmonic oscillator
where it will be demonstrated explicitly that this new method is capable of
including single-boson images with the correct asymptotic energy.

%%%%%%%%%%%%%%%%%%%%%%%%%%%%%%%%%%%%%%%%%%%%%%%%%%%%%%%%%%%%%%%%%%%%%%%
%
\section{Extended Holstein-Primakoff Mapping}
%
%%%%%%%%%%%%%%%%%%%%%%%%%%%%%%%%%%%%%%%%%%%%%%%%%%%%%%%%%%%%%%%%%%%%%%%

As a starting point let us consider a system with two types of bosonic creation and
annihilation operators: $a^+$, $a$, and $b^+$, $b$. 
Pairing these in all possible ways leads to ten group generators  
of the non-compact $Sp(4)$ group. 
The pairs $a^+a$, $aa$, $a^+a^+$ and analogously the pairs of  $b$-operators form two
commuting $Sp(2)$ subgroups.
The number conserving bilinears $a^+a$, $b^+b$, $a^+b$ and $ab^+$ span a closed
$U(2)$ algebra. There remain the bilinears $a^+b^+$ and $ab$ which do not belong to
any non-trivial subgroup of $Sp(4)$. \\
 Our goal will be to first set up the boson images 
of the ten group generators, replacing in the end the $b$-operators by c-numbers
(the condensate). This will lead us to the boson image of the
semidirect product group $Sp(2)\otimes N(1)$
made up of the elements $a^+a$, $aa$, $a^+a^+$, $a$, $a^+$, and $1_d$
respectively.
The latter is the desired system because it involves even and odd
number of boson operators.\\   
We will follow earlier work for interacting Fermions by 
Evans and Kraus  \cite{EK},  
Klein, Cohen, Li, Rafelski, and Rafelski \cite{KRR,KCL} in which 
a mapping for the ten generators of the $SO(5)$ group was derived. 
Especially, use is made of the work of the latter group of authors to derive
this time the mapping of the ten generators of the non-compact $Sp(4)$ group mentioned
above.
Since there is no room to go into details (which will be presented elsewhere) 
we essentially will only give the result here.\\
One first realizes that the six generators of the two
commuting $Sp(2)$ algebras can be mapped via the usual HPM.
The difficult task lies 
in finding an adequate mapping for the generators $a^+b^+$ and $ab$,
which allows one to close of the full $Sp(4)$ algebra. 
The reader is invited to consult reference \cite{KRR}
for a similar derivation.
Introducing a set of three new bosonic operators $\alpha$,
$A_1$, and  $A_2$, one can show that the net result for the complete
mapping reads
\begin{eqnarray}
 (a^+ a^+)_{I} &=& A_1^+ \sqrt{ 2 \,+\, 4(n_1+ m)} ~,
\quad\quad  (a a)_{I} = \left((a^+ a^+)_{I}\right)^+~,\quad\quad 
   (a^+ a)_{I} =  2n_1+ m~, 
\nonumber\\
(b^+ b^+)_{I} &=& A_2^+ \sqrt{ 2 \,+\, 4(n_2+ m)} ~,
\quad\quad  (b b)_{I} = \left((b^+ b^+)_{I}\right)^+~,\quad\quad 
(b^+ b)_{I} =  2n_2+ m~, 
\nonumber\\
  (a^+ b^+)_{I} &=& 
\alpha^+ \,\sqrt{ 2 \,+\, 4(n_1+ m)}\sqrt{ 2 \,+\, 4(n_2+ m)} \,\Phi(m)\,\,
+\,\, 4\, \Phi(m) A^+_2 A^+_1 \alpha ~, \nonumber\\ 
 (a b)_{I} &=& \left((a^+ b^+)_{I}\right)^+~, 
\nonumber\\
  (a^+ b)_{I} &=& \frac{1}{2} \, \left[\, (b b)_{I} \,,\, (a^+ b^+)_{I}\,
\right]\,,\quad\quad\quad (b^+ a)_{I} =\left((a^+ b)_{I} \right)^+~,
\label{eq1}
\end{eqnarray}
where $n_1, n_2$, and  $m$ are  occupation number operators defined by 
\begin{equation}
m = \alpha^+\alpha~,\quad\quad\quad\quad n_i = A_i^+A_i \quad\quad (i=1,2)~. 
\label{eq2}
\end{equation}
The $+$-sign in the  Holstein-Primakoff square-root
indicates the non-compact character of the group at hand. 
Finally, the function $\Phi$ is given by
\begin{equation}
 \Phi(m) \,=\, \left[\frac{r\,+\,m^2}
{4(m\,+\,1)(2 m\,+\,1)(2 m\,-\,1)}\right]^{\frac{1}{2}}~,
\label{eq3}
\end{equation}
where $r$ is a constant which is fixed using physical conditions as will
be discussed in the next section.\\
These results constitute only an intermediate step towards our final goal.       
As stated in the introduction, one wishes to extend the usual HPM
for boson pairs, in such a way as to allow the mapping of single bosons, as well.
In other words, and following the original Belyaev-Zelevinsky
approach, one needs to achieve a realization of the following algebra
\begin{eqnarray} 
\left[  a a \,,\, a^+ a^+\right]
\,&=&\, 2\,  \,\,+\,\,
 4\, a^+ a~, \nonumber\\
\left[  a a\,,\,  a^+  a\right] \,&=&\, 2\,  a  a~,
\nonumber\\
\left[  a \,,\, a^+  a^+\right]\,&=&\, 2\, a^+ ~, \nonumber\\
\left[  a \,,\, 
a^+  a\right]\,&=&\,  a~,
\label{eq4}
\end{eqnarray}
where all other possible commutators are assumed but not explicitly shown here.
This is nothing but the algebra of the semidirect product group 
$Sp(2)\otimes N(1)$. The first two commutation relations in Eq.~(\ref{eq4}) 
remind us of the $Sp(2)$ algebra, and as such, one can propose the bosonic HPM
as a second realization for it. Here again, the difficulty lies in finding
an adequate mapping for the single bosons so as to close the algebra above.
A way out is to notice that, by considering the limit in which the operator $b$ 
and $b^+$ are transformed into the identity operator, one can ultimately contract
the whole $Sp(4)$ group to a non-isomorphic semidirect group 
$Sp(2)\otimes N(1)$. This singular transformation, which can be thought of as a
contraction \`a la In\"{o}n\"{u}-Wigner or Saletan \cite{INWI,SAL}, gives a clear 
hint on how to proceed with the desired extension. 
Indeed, the single bosons can be deduced from the contraction 
of the generators $a^+b^+$, $ab$, $a^+b$, and $ab^+$ down to the generators
$a$ and $a^+$. With this intuitive picture in
mind, one can show that the following mapping for the five relevant generators 
constitutes a realization of the algebra in Eq.~(\ref{eq4}).
\begin{eqnarray} 
(a a)_I\,&=&\, \sqrt{2\,+\,4(n_1 + m)}\, A_1~,\quad\quad\quad\quad  
(a^+ a^+)_I \,=\, ( a  a)_I^+~,
\quad\quad\quad\quad 
(a^+ a)_I\, =\, 2\,n_1 \,+\, m ~,
\nonumber\\
(a)_I\,&=&\, \sqrt{2\,+\,4(n_1 + m)} \,\Gamma_1(m) \,\alpha \,\,+\,\, 
2\, \alpha^+\, A_1 \,\Gamma_1(m)~,
\quad\quad\quad\quad\quad
(a^+)_I \,=\, (a)_I^+~,
\label{eq5}
\end{eqnarray}
where the occupation number operators are, as before, 
given by $n_1 = A_1^+A_1 \,,\quad m =  \alpha^+\alpha $, while the function 
$\Gamma_1$ reads
\begin{equation}
 \Gamma_1(m )=
\left[\frac{z_1 + m^2 }{2(m+1)(2m+1)(2m-1)}\right]^{\frac{1}{2}}~.
\label{eq6}
\end{equation}
Here too $z_1$ is a constant which will be fixed by using physical conditions
as will be explained in the next section. It is  straightforward to verify, 
through a direct evaluation of the commutators in Eq.~(\ref{eq4}), that this 
is indeed a proper realization. 
This completes our considerations concerning the mapping.
In the next section, the formalism will be applied to 
the interesting case of $N$ oscillators and used to develop the
$1/N$-expansion.

%=====================================================================
%
\section{The $O(N+1)$ Anharmonic Oscillator}
%
%======================================================================

As an application, let us consider
the anharmonic oscillator with an $O(N+1)$ symmetry broken down to $O(N)$.
The properly scaled Hamiltonian of the system is given by
\begin{equation}
 H  =  \frac{ {\vec P}_{\pi}^2}{2} +\frac{ P_{\sigma}^2}{2}  
 + \frac{\omega^2}{2} \left[ {\vec X}_{\pi}^2 + X_{\sigma}^2 \right]
 + \frac{g}{N} \left[ {\vec X}_{\pi}^2 + X_{\sigma}^2 \right]^2
   - \sqrt{N}\eta X_{\sigma}~.
\label{eq7}
\end{equation}
Here we have considered an explicit $(\eta \neq 0)$ and a 
spontaneous $(\ave{X_{\sigma}} \neq 0)$  symmetry breaking along the
$X_{\sigma}$ mode.    
The variables ${\vec X}_{\pi},  X_{\sigma}$ and their conjugate momenta  
${\vec P}_{\pi},  P_{\sigma}$ are expressed in second quantization as
\begin{eqnarray}
 {\vec X}_{\pi} \,&=&\, \frac{1}{\sqrt{2\omega}}({\vec a} + {\vec a}^+)~,\quad\quad\quad\quad
 {\vec P}_{\pi} \,=\, i \sqrt{\frac{\omega}{2}}\,
 ({\vec a}^+ - {\vec a})~,  \nonumber \\
 X_{\sigma} \,&=&\, \frac{1}{\sqrt{2{\cal E}_{\sigma}}}(b + b^+)~,\quad\quad\quad\quad
 P_{\sigma} \,=\, i \sqrt{\frac{{\cal E}_{\sigma}}{2}}\,(b^+ - b)~.  
\label{eq8}
\end{eqnarray} 
The  frequency, ${\cal E}_{\sigma}$, of the mode $X_{\sigma}$  will be fixed later.
The subscripts $\pi$ and $\sigma$ are used in analogy with the 
linear $\sigma$-model in QFT, where these modes represent the pion-
and sigma fields respectively.\\   
To sort-out the dynamics according to the $1/N$-expansion, one needs to
adapt the mapping derived in the previous section to the situation 
of $N$ oscillators. This can be done in a straightforward way.
It can be shown that the mapping in this case takes the form
\begin{eqnarray}
({\vec a}{\vec a})_I \,&=&\, \sqrt{2N\,+\,4(n_1 +m)}\, A_1~,
\quad\quad\quad\quad
({\vec a}^+{\vec a})_I \,=\, 2\,n_1 \,+\, m ~,
\quad\quad\quad\quad
({\vec a}^+{\vec a}^+ )_I \,=\, ({\vec a}{\vec a})_I^+~,
\nonumber\\
 (a_i)_I \,&=&\, \sqrt{2N\,+\,4(n_1+m)} \,\Gamma_N(m) \,\alpha_i \,\,+\,\, 
2\, \alpha_i^+\, A_1 \,\Gamma_N(m)~,\quad\quad\quad\quad\quad\quad\quad
(a_i^+)_I = (a_i)_I^+~,
\label{eq9}
\end{eqnarray}
where $N$ is an integer,  $n_1 = A_1^+A_1$, and 
$ m = \sum_i \alpha_i^+\alpha_i $,
while $\Gamma_N$ is a generalization of 
the $\Gamma_1$ function, of the last section  
to the case of $N$ oscillators. It reads
\begin{equation}
 \Gamma_N(m )=
\left[\frac{z_N + m^2 +m(N-1)}
{2(m+1)(2m+N)(2m+N-2)}\right]^{\frac{1}{2}}~.
\label{eq10}
\end{equation}
The constant $z_N$ will be fixed below. One can also easily verify  that this 
mapping leads to a realization of the following algebra: 
\begin{eqnarray} 
\left[ \left({\vec a}{\vec a}\right)\,,\, 
\left({\vec a}^+ {\vec a}^+\right)\right]
\,&=&\, 2\, N \,\,+\,\,
 4\, \left({\vec a}^+{\vec a}\right)~, \nonumber\\
\left[ \left({\vec a}{\vec a}\right)\,,\, 
\left({\vec a}^+ {\vec a}\right)\right]
\,&=&\, 2\, \left({\vec a}{\vec a}\right)~.
\nonumber\\ 
\left[  a_i \,,\, 
\left({\vec a}^+ {\vec a}^+\right)\right]
\,&=&\, 2\, a^+_i ~, \nonumber\\
\left[  a_i \,,\, 
\left({\vec a}^+ {\vec a}\right)\right]
\,&=&\,  a_i~.
\label{eq11}
\end{eqnarray}
For a finite $N$, the $O(N+1)$ anharmonic oscillator is purely
quantum mechanical. For an infinite number of degrees of freedom 
$N \rightarrow \infty$, on the other hand, it can be used to mimic
the quantum-field situation of the breaking and restoration 
of a continuous symmetry. \\
Using the mapping in Eq.~(\ref{eq9}), one can expand   
the Hamiltonian of the system in powers of the operators 
$A$, $\alpha$, $b$ and their hermitian conjugates. One then arrives at an
expansion of the form
$H \,=\,{\cal H}^{(0)}\,+\,{\cal H}^{(1)}\,+\,{\cal H}^{(2)}
\,+\,{\cal H}^{(3)}\,+\,{\cal H}^{(4)}\,+...$, where the superscripts indicate
powers of operators without normal ordering.
This expansion is in fact not unique and therefore  
the preservation of the symmetries is not necessarily guaranteed. A  more useful 
approach consists in organizing the expansion in powers of the parameter $N$, 
such that~$H \,\,=\,\, N H_0\,+\,\sqrt{N}H_1\,+\,
H_2\,+\frac{1}{\sqrt{N}}\,H_3\,+\,\frac{1}{N} H_4\,+...$\\
This is possible if one chooses a coherent state as the variational ground state 
for the model
\begin{equation}
\ket{ \psi} = \exp\left[  \ave{A_1} \,A_1^+ \,\, + \,\, 
 \ave{b}\, b^+ \right] \ket{ 0}.
\label{eq12}
\end{equation}
This trial vacuum state must accommodate two condensates respectively 
for the $X_{\sigma}$ mode
and the newly introduced boson $A_1$ (see ref. \cite{ASW} for
details). The mode $\alpha$, on the other hand, is not allowed to condense.    
The ground-state energy, 
$N H_0 = \frac{\bra{\psi} H\ket{\psi}}{\ave{\psi | \psi}}$~,
calculated on the coherent state, takes 
the following form
\begin{equation}
H_0 \,=\, 
\frac{\omega}{2}\left(2d^2+1\right) +
\frac{g s^2}{ \omega} 
\left(d + \sqrt{1 +  d^2}\right)^2 + 
\frac{g}{4 \omega^2} \left( d + \sqrt{1 +  d^2}\right)^4
+ \frac{\omega^2 s}{2} + g s^4 - \eta s~,
\label{eq13}
\end{equation}
where we have introduced for convenience the rescaled condensates  
$ s = \frac{1}{\sqrt{N}}\,\ave{X_{\sigma}}~,
 d = \sqrt{\frac{2}{N}} \,\ave{A}$.\\
The coherent ground state is fully determined by requiring that the 
values taken by the two condensates above lead to the minimum of $H_0$. 
The minimization procedure with respect to $s$ and $d$  
gives the following two coupled BCS equations 
\begin{eqnarray}
\omega^2 + 4 g s^2 + \frac{2g}{ \omega} \left(d+ \sqrt{1 +  d^2}\right)^2
&=& \frac{\eta}{s}~, 
\nonumber\\
2 \omega d \sqrt{1 + d^2} \,+\,\left(d+ \sqrt{1 +  d^2}\right)^2 \Delta 
&=& 0~,
\label{eq14}
\end{eqnarray}
where $\Delta \,=\, \frac{2 g s^2}{\omega} \,+\, \frac{g}{\omega^2} 
\left(d + \sqrt{ 1 + d^2} \right)^2$ is  the gap parameter .\\
To gather the full dynamics of the leading order in the $1/N$-expansion one needs
to generate the terms $H_1$ and $H_2$ of the Hamiltonian. 
This can be done by using the parameter differentiation techniques 
(see ref. \cite{ASW} for details). The net result for both $H_1$ and $H_2$ 
then reads:
\begin{eqnarray}
H_1 &=& 
\frac{1}{\sqrt{2}} \left[ 2 \omega d + 
\frac{\left(d + \sqrt{1 + d^2}\right)^2}
{\sqrt{1 + d^2}} \Delta \right] \left( {\tilde A}_1 + {\tilde A}_1^+\right)
+  \left[ \frac{2 g s}{\omega}
\left( d + \sqrt{ 1 + d^2}\right)^2 + \omega^2 s + 4 g s^3 - \eta
\right] \frac{(\beta^+ + \beta)}{\sqrt{2 {\cal E}_{\sigma}}} 
\nonumber\\
H_2&=&  {\cal H}_0 \,+\,
{\cal E}_{\sigma} \beta^+ \beta \,+\,\left[ \omega + \Delta + \frac{\Delta d}
{\sqrt{ 1 + d^2}} \right] m + 
\left[ 2 \omega + 2 \Delta + \frac{\Delta d}
{\sqrt{ 1 + d^2}}  \right]{\tilde n}_1  \nonumber\\
&+&\left({\tilde A}_1 + {\tilde A}_1^+ \right)^2 
\left[
 \frac{\Delta d\left(2+d^2\right)}
 {4 \sqrt{\left( 1 + d^2\right)^3}}
 + \frac{g}{2 \omega^2}\frac{\left( d + \sqrt{ 1 + d^2}\right)^4}
 {1 + d^2} \right] + 2g s 
 \frac{(\beta^+ + \beta)({\tilde A}_1 + {\tilde A}_1^+ )}
 {\omega \sqrt{{\cal E}_{\sigma}}}   
 \frac{\left( d + \sqrt{ 1 + d^2}\right)^2}
 {\sqrt{ 1 + d^2}}~.  \nonumber\\
\label{eq15}
\end{eqnarray}
Here, ${\cal H}_0$ is a constant. Since one is not 
particularly interested in the ground-state energy, the latter will not be
further specified. 
The shifted operators ${\tilde A}_1 = A_1 - \ave{A_1}$,  
$\beta = b - \ave{b} $, and $ {\tilde n}_1 = {\tilde A}_1^+ {\tilde A}_1$
annihilate the coherent state $\ket{\Psi}$. 
The frequency ${\cal E}_{\sigma}$ of the $X_{\sigma}$ mode    
is fixed such that the bilinear
part of $H_2$ in the $\beta$ operators is diagonal. 
It is purely of perturbative character, and the frequency is explicitly given by 
\begin{equation}
{\cal E}_{\sigma}^2 \,=\,
 \omega^2 + 12 g s^2 + \frac{2g}{ \omega} \left(d+ \sqrt{1 +  d^2}\right)^2~.  
\label{eq16}
\end{equation}
Using the gap equations (\ref{eq14}) and the following easily verifiable identities
\begin{equation}
\Delta \,=\,-2d\sqrt{1 + d^2}\left[\omega \left( d - \sqrt{1 + d^2}\right)^2
\right],\quad\quad
\omega + \Delta \,=\,(1 + 2d^2)\left[\omega \left( d - \sqrt{1 + d^2}\right)^2
\right]~,
\label{eq17}
\end{equation}
one can establish that $H_1$  vanishes at the minimum. 
From $H_2$, 
and more precisely from the coefficient of $m=\sum_{i=1}^N \alpha_i^+ \alpha_i $,
one can deduce the existence of $N$ uncoupled modes.
The common frequency of these modes is denoted by ${\cal E}_{\pi}$ and is given
by
\begin{equation}
{\cal E}_{\pi}^2 \,=\,\omega^2 \left( d - \sqrt{1 + d^2}\right)^4\,=\,
\omega^2 + 4 g s^2 + \frac{2g}{ {\cal E}_{\pi}} \,=\,\frac{\eta}{s}~.
\label{eq18}
\end{equation}
These $N$ modes are our first asymptotic states. Furthermore,
it can be easily verified 
that they have Goldstone character. In other words, their frequency vanishes 
in the exact symmetry limit $(\eta = 0)$, 
and for a finite condensate $(s\neq 0)$.\\
It is evident
from the ansatz above that the model suffers 
from infrared divergences. However, since it is used for 
demonstration purposes only, we choose to disregard this difficulty here.  
Clearly the new and important result that has been obtained 
shows up in the fact that the proposed mapping provides
asymptotic states in the ideal Fock-space which correspond
 to the images of the single bosons. 
It should be stressed that this is a non trivial-finding which,
as shown above,   is a direct consequence 
of the extended HPM. It reproduces the result anticipated 
in the introduction,
in clear departure from the HPM for boson pairs \cite{ASW} 
and in  accordance with 
the Goldstone theorem.\\
So far, only the mapping of the bilinears in Eq.~(\ref{eq9}) was involved in 
expanding the Hamiltonian. The single-boson part of the mapping, on the other
hand, was not directly used. The latter enters, however, in the definition of
the two-point function $\bra{\Psi} T X_{\pi,i}(t) X_{\pi,i}(t') \ket{\Psi}$,
where $\ket{\Psi}$ is the coherent ground state.
To leading order in $1/N$ and after a Fourier transform one obtains
\begin{equation}
D_{\pi,\,ij}(s) \,=\, \int dt \,e^{i\sqrt{s}(t-t')}\, 
\bra{\Psi} T X_{\pi,i}(t) X_{\pi,j}(t') \ket{\Psi} \,=\, \delta_{ij}
\frac{2\,N\, \Gamma_N^2(0)}{s \,-\,{\cal E}_{\pi}^2 \,+\,i\eta}~.
\end{equation}
The fact that the residue at the pole has to be~ $2\,N\, \Gamma_N^2(0)=1$,
 leads to  $z_N = N-2$.\\  
Besides the Goldstone modes there also exist other
excitations. They can be made explicit in
diagonalizing the remaining part of $H_2$. 
This is a straightforward procedure which can be found in
\cite{ASW}. In short, since the non-diagonal part of $H_2$ is at most bilinear
in the operators ${\tilde A}_1, {\tilde A}_1^+, \beta, \beta^+$, a generalized 
Bogoliubov rotation of the type
\begin{equation}
Q_{\nu}^+ \,=\, X_{\nu} \beta^+ \,-\,Y_{\nu} \beta \,+\, 
U_{\nu} {\tilde A}_1^+ \,-\, V_{\nu} {\tilde A}_1~,
\label{eq20}
\end{equation}
can be performed and leads to uncoupled modes at the minimum of the
action. The diagonalization is done by recalling the usual Rowe equations of motion
\cite{RS80}      
\begin{equation}
\bra{RPA} \left[ \delta Q_{\nu}\,,\, \left[H_2 \,,\, Q_{\nu}^+ \right]\right]\ket{RPA}
 \,=\, \Omega_{\nu} 
\bra{RPA} \left[ \delta Q_{\nu}\,,\, Q_{\nu}^+ \right]\ket{RPA}~,
\label{eq21}
\end{equation}
where $\ket{RPA}$, the full ground state of the theory at this order,
is a random-phase approximation (RPA) ground state, defined by  
$Q_{\nu}\ket{RPA} =0$.
The Hamiltonian can then be written in the RPA phonon basis, 
$\ket{\nu}= Q^+_{\nu}\ket{RPA}$, as follows
\begin{equation}
H \,=\, N H_0  \,+\,  E_{RPA}\,+\,{\cal E}_{\pi} 
\sum_{i=1}^N \alpha_i^+\alpha_i 
\,+\,  \sum_{\nu = {\pm}1,{\pm}2} \Omega_{\nu} Q_{\nu}^+ Q_{\nu} \,+\, 
{\cal O}(N^{-\frac{1}{2}})~. 
\label{eq22}
\end{equation}
and contains three terms of order $(\sqrt{N})^2, (\sqrt{N})^1, (\sqrt{N})^0,$
respectively. The coefficient of the $\sqrt{N}$ term vanishes.
The contribution $ E_{RPA} =\ave{RPA | H_2 | RPA}$ is the RPA correction to the 
ground-state energy and will not be given explicitly here.
The frequencies $\Omega_{\nu}$  are solutions of the 
characteristic equation of the RPA eigenvalues problem and given by
\begin{equation}
\Omega_{\nu}^2 \,=\, \frac{\eta}{s} \,+\,  
\frac{8 g s^2}{1\,\,-\,\, \frac{4g}{{\cal E}_{\pi}}
\frac{1}{\Omega_{\nu}^2 \,-\,4 {\cal E}_{\pi}^2}}~.
\label{eq23}
\end{equation}
In the exact symmetry limit $(\eta = 0)$, there exist a pair of zero-energy 
solutions among the four RPA eigenvalues
which correspond to two uncorrelated Goldstone modes
\footnote{ Here again, we disregard the infrared problem 
since the model is only used for demonstration purposes.
The reader is referred to \cite{ASW}
for a thorough study of these questions in four space-time dimensions.}.
This point is not the mean purpose to the present note and  
therefore will not be  discussed further. The reader may consider looking
into ref.\cite{ASW} for a complete treatment of this question. \\
We therefore see that the Hamiltonian in  Eq.~(\ref{eq22}) is the same as in 
\cite{ASW}, however, augmented by the 'single-pion' term 
$\sum_{i=1}^N \alpha_i^+\alpha_i$. This extra term arises necessarily in our
approach where single bosons and pairs of bosons are treated on the same footing.
In \cite{ASW} the single boson state has been treated on a heuristic level by 
neglecting exchange contributions to the self energy. So implicitly, this 
amounts to the same as using (\ref{eq22})
at the order considered. The present systematic scheme puts the treatment
of ref.~\cite{ASW} on a firm theoretical ground.

%%%%%%%%%%%%%%%%%%%%%%%%%%%%%%%%%%%%%%%%%%%%%%%%%%%%%%%%%%%%%%%%%%%%%%%%
%
\section{Conclusion}
%
%%%%%%%%%%%%%%%%%%%%%%%%%%%%%%%%%%%%%%%%%%%%%%%%%%%%%%%%%%%%%%%%%%%%%%%%%
In this paper we have extended previous work on the 
Holstein-Primakoff boson expansion for boson pairs applied 
to a relativistic field theory of interacting bosons \cite{ASW}. The
aim was to treat simultaneously single bosons and pairs of bosons
which is necessary to unambiguously define the two-point function
for the Goldstone mode and to extend the formalism to finite temperature.\\   
The mapping was applied to the anharmonic oscillator with broken $O(N+1)$-symmetry. 
It was explicitly shown that the extension to accommodate single bosons indeed renders, 
to leading order of the $1/N$-expansion, $N$ uncoupled Goldstone- as well 
as RPA phonon modes.
This result is novel and inaccessible to the  bosonic 
Holstein-Primakoff mapping for boson pairs.  The latter 
is only able to provide RPA phonon modes as previously shown in
ref.~\cite{ASW}.
  The full power of the formalism will reveal
itself in working out the next-to-leading order 
of the $1/N$-expansion by providing an unambiguous
computation of all $n$-point functions.    
It also allows for a natural and straightforward extension to finite temperature.
These two points will be discussed in a forthcoming publication.

\vspace*{0.5cm}
\begin{center}
\underline{\bf \large Acknowledgments:}
\end{center}
I would like to thank G. Chanfray, P. Schuck and  J. Wambach for 
the fruitful collaboration and for their 
continuous support. 
I also would like  to thank P. Schuck
and J. Wambach for discussions, for their interest in this work, 
and for their comments on the manuscript. 
Finally, I would like to  thank the Gesellschaft f\"ur Schwerionenforschung (GSI) Darmstadt
for the financial support.

\end{document}